\documentclass[aps,twocolumn,preprintnumbers]{revtex4}

\usepackage{graphicx}  
\usepackage{subfigure}
\usepackage{multirow}

\usepackage{fancyhdr}
\usepackage{longtable}
\usepackage{parskip}
\usepackage[T1]{fontenc}
\usepackage{dcolumn}   
\usepackage[dvipsnames]{xcolor}
\usepackage{bm}        
\usepackage[normalem]{ulem} 
\usepackage{amsfonts}  
\usepackage{amsmath}   
\usepackage{amssymb}   
\usepackage{hyperref}
\usepackage[normalem]{ulem}

\newcommand{\pwisein}{\left\{ \begin{array}{ll}}
\newcommand{\pwiseout}{\end{array}\right.}

\definecolor{darkgreen}{RGB}{0,100,0}
\begin{document}

\title{Kerr nonlinearity and three-wave mixing in superconducting resonators hosting Al-InAs weak links}

\author{Vittorio Buccheri$^1$\footnote{buccheri@chalmers.se}, Ivo P. C. Cools$^1$, Nermin Trnjanin$^1$, Ankit Khola$^1$, Oleg Shvetsov$^1$, Thomas Kanne$^2$, Jesper Nyg{\aa}rd$^2$, Attila Geresdi$^1$, Simone Gasparinetti$^1$\footnote{simoneg@chalmers.se}}

\affiliation {\it $^1$Department of Microtechnology and Nanoscience, Chalmers University of Technology, 41296 Gothemburg, Sweden.\\ $^2$Center for Quantum Devices, Niels Bohr Institute, University of Copenhagen, Universitetsparken 5, DK-2100 Copenhagen, Denmark.}

\begin{abstract}  
Nonlinear microwave resonators are a versatile tool in quantum information processing, enabling parametric amplification, continuous variable quantum computing, and engineered mode interactions. Many of these applications especially benefit from cubic nonlinearities enabling three-wave mixing; at the same time, they are limited by quartic nonlinearities giving rise to undesired Kerr effects. A recurrent challenge is therefore to engineer resonators with a finite cubic nonlinearity while suppressing quartic terms. Here, we investigate a superconducting resonator hosting two weak links fabricated from an aluminum-capped indium arsenide nanowire. We characterize the Kerr nonlinearity as a function of magnetic flux and gate bias, showing that it can be tuned to zero with either control parameter. Furthermore, we experimentally demonstrate three-wave mixing in a semiconductor-superconductor hybrid device, establishing nonzero cubic nonlinearity. An effective model based on Andreev bound states qualitatively captures the observed trends. Our results validate semiconductor-superconductor hybrid devices as a promising platform for tunable nonlinear superconducting circuits, with applications in parametric amplification, quantum control of bosonic modes, and engineering  interactions between microwave modes.
\end{abstract}

\maketitle 

\begin{figure*} [ht!]
        \begin{center}
                \includegraphics [width=1\textwidth]{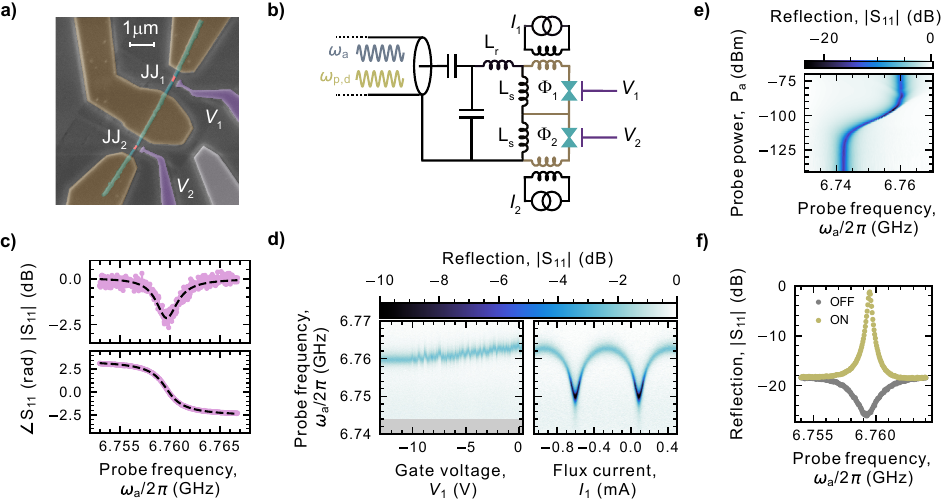}
        \end{center}
        \caption{Device and characterization. \textbf{a)} False-colored scanning electron micrograph showing two InAs junctions (red), $\rm{JJ}_1$ and $\rm{JJ}_2$, defined by selectively etching a 250-nm section of the Al shell (green) covering the nanowire. The nanowire is embedded in the resonator structure (made of NbTiN, not shown) via Ti-Al contacts (yellow) obtained from a standard lift-off process, together with the gate lines $V_1$ and $V_2$ (purple), and a ground screening line (light gray) separating the two gates. \textbf{b)} Simplified circuit diagram. An LC resonator with total inductance $L_{\rm r}+ 2L_{\rm s}$ shares two equal inductive sections $L_s$ with two gate-tunable weak links arranged in superconducting loops. The magnetic fluxes through the loops, $\Phi_1$ and $\Phi_2$, are controlled by the currents $I_1$ and $I_2$. The resonator is capacitively coupled to a feedline used for both probe tone, $\omega_{\rm a}$, and pump(drive) tone, $\omega_{\rm p(d)}$. \textbf{c)} Magnitude (top) and phase (bottom) of the reflection coefficient, S$_{11}$, in a gate configuration where both junctions are pinched-off. Black dashed lines represent a circle-fit model \cite{Probst} with resonant frequency $\omega^0_{\rm{r}}/2\pi=6.759$ GHz, internal quality factor $Q^0_{\rm{i}}=2.1\times10^4$ and external quality factor $Q^0_{\rm{c}}=2.5\times10^3$. \textbf{d)} Single-tone spectroscopy showing S$_{11}$ as a function of $\omega_{\rm a}$ and $V_1$ for $I_1=-0.25$ mA (left), displaying the activation of JJ$_1$ for $V_1>-10$ V, and as a function of $\omega_{\rm a}$ and $I_1$ for $V_1=-2$ V (right), displaying a periodic flux modulation. In both cases JJ$_2$ is pinched-off ($V_2<-8$~V). \textbf{e)} Single-tone spectroscopy versus probe power $P_{\rm{a}}$ with both junctions active ($V_{1}=V_2=-1$~V). \textbf{f)} Low-power resonator spectra measured with (yellow) and without (gray) an additional pump tone at $\omega_{\rm p}/2\pi=13.518$ GHz and power $-56$ dBm with both junctions active ($V_{1(2)}=0.5(0.3)$~V).}
        \label{fig1}
\end{figure*}

Superconducting microwave resonators with engineered nonlinearities are a key component in quantum information processing. Their nonlinear response enables a broad range of applications, including parametric amplification \cite{Yurke, Yamamoto, Beltran, Zorin, Frattini_1, Frattini_2, Sivak}, bosonic-mode quantum computing \cite{Puri, Lescanne, Ma, Iyama, Eriksson}, and both linear \cite{Chapman, Maiti} and nonlinear \cite{Ye} coupling schemes.

Many of these applications rely on the cubic nonlinearity, which enables three-wave mixing for parametric amplification \cite{Zorin, Frattini_1, Frattini_2, Sivak}, non-Gaussian interactions \cite{Eriksson}, and beamsplitter coupling between bosonic modes \cite{Chapman}. By contrast, the Kerr nonlinearity is often detrimental, limiting the dynamic range of the amplifiers and degrading the fidelity of bosonic-mode operations. Engineering resonators with a nonzero cubic nonlinearity while suppressing the Kerr term is therefore a recurrent challenge in superconducting quantum circuits.

Nonlinearity is introduced into superconducting resonators either through the current-dependent kinetic inductance of a superconducting film \cite{Tholen, Malnou, Splitthoff, Buccheri} or through Josephson-junction-based elements. Devices such as the radio-frequency superconducting quantum interference device (rf-SQUID) \cite{Zorin}, the superconducting nonlinear asymmetric inductive element (SNAIL) \cite{Frattini_1}, the asymmetrically threaded SQUID (ATS) \cite{Lescanne}, and the linear inductive coupler (LINC) \cite{Maiti} exploit the nonlinear Josephson potential, whose effective shape can be engineered by arranging one or more junctions in superconducting loops. Appropriate choices of the junction parameters and external magnetic flux then allow independent control of the cubic and quartic nonlinearities. 

Another strategy for realizing tunable nonlinearity relies on semiconducting weak links. In the short-junction limit, a single conducting-channel semiconducting weak link hosts a spin-degenerate pair of Andreev bound states (ABSs) with potential $\pm U_{\rm{ABS}}(\varphi)=\pm\Delta\sqrt{1-\tau\sin^2\left(\varphi/2\right)}$, where $\Delta$ is the superconducting gap of the junction leads and $\tau$ is the junction transparency \cite{Beenakker, Park, Metzger}. In contrast to tunnel junction-based technologies, $\tau$ can be tuned via electrostatic gating, providing an additional control parameter alongside magnetic flux. 

The ABSs phenomenology has been widely investigated, including coherent manipulation of both charge \cite{Janvier, Hays_1, Zellekens} and spin \cite{Tosi, Hays_2, Hays_3, Bargerbos, Vidal, Shvetsov} degrees of freedom. However, the dependence of ABS-induced nonlinearities in superconducting resonators on the electrostatic gating and external magnetic flux has not yet been investigated experimentally. This behavior has only been discussed theoretically \cite{Schrade}, while experimental studies have been limited to power-dependent state discrimination \cite{Tosi_2}. Moreover, although parametric amplification based on four-wave mixing has already been demonstrated in aluminum (Al)-indium arsenide (InAs) hybrid devices, both in planar heterostructures \cite{Phan, Hao} and nanowire platforms \cite{Zenou}, the three-wave-mixing amplification has not yet been reported. 

In this work, we investigate a superconducting resonator incorporating two Al-InAs weak links. We study how these weak links induce both a linear and nonlinear response in the resonator, including the resonance-frequency shift, the Kerr nonlinearity, and the first realization of three-wave mixing parametric amplification. We show that these properties can be tuned through the electrostatic gates controlling the junction transparencies and through the external magnetic flux setting the superconducting phase difference across the weak links. Furthermore, we demonstrate that the observed behavior is qualitatively captured by an effective model based on the ABSs phenomenology. Our results establish hybrid semiconductor-superconductor weak links as versatile nonlinear elements for applications in parametric amplification, bosonic-mode manipulation, and engineered coupling schemes.

The core of our device consists of an Al-capped InAs nanowire \cite{Krogstrup} hosting two Josephson junctions [see Fig.~\ref{fig1}a)]. The junctions are approximately 250 nm long and separated by 4 $\mu$m. The nanowire is embedded in a niobium-titanium-nitride (NbTiN) resonator structure [see Supplementary for details on the exact resonator topology and coupling scheme]. An equivalent circuit representation of the device consists of an LC resonator where two portions, $L_{\rm{s}}$, of the total inductance are shared with the junctions, JJ$_1$ and JJ$_2$, forming a double rf-SQUID configuration [see Fig.~\ref{fig1}b)]. The resonator is capacitively coupled to a feedline and is probed in reflection; the same feedline is used for drive and pump tones. The electrochemical potentials of the semiconducting junctions are controlled by two gate lines, while two flux lines set the superconducting phase across them. The resonator used in this work has been discussed in \cite{Cools}, and similar devices designed for single junction studies have been employed in \cite{Hays_1, Hays_2, Hays_3, Shvetsov}. We perform our experiments in a dilution cryostat down to 10 mK. To characterize the device, we compensate for the crosstalk between the flux and gate lines, enabling independent control of the magnetic flux and the electrostatic potential of each junction.

The gate voltages define four distinct operating regimes. When both gate voltages are below their respective pinch-off thresholds, $\tilde{V}_1=-10$ V and $\tilde{V}_2=-8$ V, the semiconducting weak links are fully depleted, suppressing transport through both junctions. Here, the junctions effectively behave as open circuits, and the resonator response is independent of both gate voltage and magnetic flux. In this regime we extract the bare resonant frequency, $\omega^0_{\rm r}$, and the internal(external) quality factor, $Q^0_{\rm{i(c)}}$ [Fig.~\ref{fig1}c)]. Increasing only one gate voltage, namely $V_1(V_2)\gg\tilde{V}_1(\tilde{V}_2)$ while keeping $V_2(V_1)<\tilde{V}_2(\tilde{V}_1)$, activates a single-junction configuration and, finally, increasing both gate voltages above pinch-off activates both junctions, bringing the device into the double-junction regime.

Activating a junction opens several conduction channels, whose number and transparency are strongly gate dependent. Consequently, the resonance frequency exhibits an irregular dependence on gate voltage [Fig.~\ref{fig1}d), left], consistent with previous observations \cite{Tosi,Shvetsov}. An active junction acquires a phase-dependent Josephson inductance resulting in the periodic modulation of the resonance frequency $\omega_{\rm r}(I_1)$, characteristic of a SQUID-like circuit [Fig.~\ref{fig1}d), right]. The resonator quality factor is also modulated, as evidenced by changes in the signal intensity. Near the maximum of $\omega_{\rm r}(I_1)$, the resonator remains overcoupled and close to its bare parameters, whereas near the minimum, the resonance shifts to lower frequencies and the losses increase, bringing the resonator close to critical coupling.

Beyond shifting the resonance frequency, both the single- and double-junction configurations introduce nonlinearities, including Kerr nonlinearity and three-wave mixing \cite{Sivak,Frattini_1}. The Kerr nonlinearity manifests as a power-dependent shift of the resonance frequency, $\delta \omega_{\rm{kerr}}(P_{\rm a})$, defined with respect to its low-power value [Fig.~\ref{fig1}e)], whereas the three-wave mixing enables energy transfer from a pump tone at $\omega_{\rm p}$ to signal and idler tones at $\omega_{\rm s}$ and $\omega_{\rm i}$ satisfying $\omega_{\rm p}=\omega_{\rm s}+\omega_{\rm i}$. In the degenerate case, $\omega_{\rm s}=\omega_{\rm i}$, yielding $\omega_{\rm p}=2\omega_{\rm s}$. By comparing two probe-frequency sweeps around $\omega_{\rm r}/2\pi=6.759$ GHz with and without a pump tone at $\omega_{\rm p}/2\pi=13.518$ GHz, we observe an on-off gain exceeding 20 dB [Fig.~\ref{fig1}f)], which demonstrates three-wave mixing in our system. From the ratio between the on and off traces we extract the on-off gain $G$, as the amplitude of the Lorentzian fitted to the gain profile (see Supplementary).

\begin{figure*} [ht!]
        \begin{center}
                \includegraphics [width=1\textwidth]{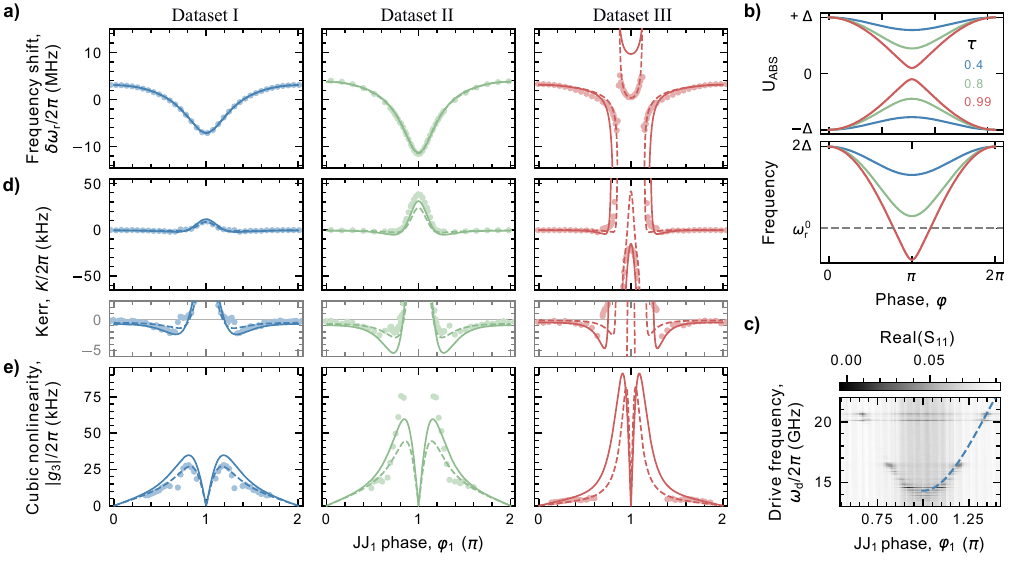}
        \end{center}
        \caption{Single-junction configuration (JJ$_1$ active, JJ$_2$ pinched-off). \textbf{a)} Measured frequency shift, $\delta \omega_{\rm r}$, as a function of the phase across JJ$_1$, $\varphi_1$, for Dataset I (left, $V_1=0.35$ V), II (center, $V_1=2.2$ V) and III (right, $V_1=1.25$ V). Solid and dashed lines reproduce single- (Eq.~\ref{df}) and double-channel (Eq.~\ref{df_2ch}) model, respectively, with the parameters given in Tab.\ref{tabS_fit_param1}. \textbf{b)} Single Andreev bound state (ABS) model for three representative transparencies $\tau$. \textit{Top}: Ground- and excited-state ABS energies versus superconducting phase, $\pm U_{\rm{ABS}}(\varphi)$. \textit{Bottom}: Corresponding ABS excitation frequency $2U_{\rm{ABS}}(\varphi)$. The horizontal dashed line indicates the bare resonator frequency $\omega^0_{\rm r}$. \textbf{c)} Two-tone spectroscopy measured around $\varphi_1=\pi$ for the same gate configuration of Dataset I, revealing a spectroscopic feature consistent with an ABS transition coupled to the resonator mode. The blue dashed line shows the expected $\omega_{\rm ABS}$, calculated using $\Delta/2\pi=16~\mathrm{GHz}$ and $\tau=0.8$. \textbf{d, e)} Kerr nonlinearity $K$ and cubic nonlinearity $|g_3|$ measured as function of $\varphi_1$ for Datasets I, II and III. For $K$ (panel d), both the full range (top row) and a zoom around $K/2\pi=0$ (bottom row) are shown. Horizontal gridlines on the bottom row of panel d highlight $K/2\pi=0$. Experimentally, $K$ and $|g_3|$ are obtained from $\delta \omega_{\rm{kerr}}$ and gain $G$ as explained in the main text and in the Supplementary. Solid and dashed lines represent single- (Eqs.~\ref{g3} and~\ref{K}) and double-channel (Eqs.~\ref{g3_2ch} and~\ref{K_2ch}) model, respectively, with the parameters given in Tab.\ref{tabS_fit_param1}.}
        \label{fig2}
\end{figure*}

We now measure the linear [$\delta \omega_{\rm r}=\omega_{\rm r}(I_1)-\omega^0_{\rm r}$] and nonlinear ($\delta \omega_{\rm{kerr}}$, $G$) response of the resonator by varying both the flux current and gate voltage. 
 
We first consider the single-junction configuration with JJ$_2$ pinched-off. After converting $I_1$ into the superconducting phase $\varphi_1$ (see Supplementary), we measure a full oscillation $\delta \omega_{\rm r}(\varphi_1)$ for three representative gate voltages, $V_1=(0.35, 2.2, 1.25)$ V, corresponding to Dataset I, II, and III, respectively [Fig.~\ref{fig2}a)]. We find that $\delta \omega_{\rm r}$ exhibits a larger dip when going from I to II and eventually shows an avoided-crossing pattern in III. 

These observations are compatible with tuning the transparency of ABS pairs. In the short-junction limit, a semiconducting weak link supporting a single conduction channel is described by a pair of ABSs with energies $\pm U_{\rm ABS}$ [Fig.~\ref{fig2}b), top] \cite{Park, Metzger}. Assuming $\hbar=1$, the corresponding excitation frequency, $\omega_{\rm ABS}=2U_{\rm ABS}$, reaches its maximum value, $2\Delta$, at $\varphi=0$ and $2\pi$, and its minimum value, $2\Delta\sqrt{1-\tau}$, at $\varphi=\pi$ [Fig.~\ref{fig2}b), bottom]. Since $\tau$ is gate-dependent, the minimum excitation frequency is tuned by the gate voltage. 
Consequently, the interaction between the resonator and the ABS transition depends strongly on both $\varphi$ and $\tau$, evolving from a strongly detuned regime to dispersive and eventually resonant coupling around $\varphi=\pi$ as the ABS transparency increases. This interpretation is further supported by two-tone spectroscopy measurements around $\varphi_1=\pi$, which reveal spectroscopic signatures consistent with ABS transitions [Fig.~\ref{fig2}c) and Fig.~\ref{figS_2tone_JJ1}]. 

We write the frequency shift $\delta\omega_{\rm{r}}$ induced by a single ABS pair as \cite{Park, Metzger}:
\begin{equation}
        \delta \omega_r = -\lambda^2c_2 + g_{\rm c}^2\left(\frac{2}{\omega_{\rm{ABS}}}-\frac{1}{\delta_-}-\frac{1}{\delta_{+}}\right),
    \label{df}
\end{equation}
where $\lambda$ is a dimensionless participation ratio, $c_2=d^2U_{\rm{ABS}}/d\varphi^2$, $\delta_{\pm}=\omega_{\rm{ABS}}\pm\omega^0_{\rm{r}}$ and $g_{\rm c}=\lambda\sqrt{1-\tau}\left(dU_{\rm{ABS}}/d\varphi\right)\tan\left(\varphi/2\right)$ is the phase-dependent coupling rate. The first term in Eq.~\ref{df} is the contribution coming from the curvature of the ABS ground state potential, which is dominant at low $\tau$, whereas the second term takes into account the exchange of virtual photons between the resonator and the ABS pair, which is dominant at high $\tau$.

The short-junction limit in InAs nanowire has been proven accurate up to $\sim100$ nm-long junctions \cite{Abay}. Even if our junctions are more than twice as long, following Refs.~\cite{Metzger, Shvetsov, Iglesias}, we use the short-junction results as an effective phenomenological description, where finite junction length effects and multichannel transport are modeled through a gate-dependent effective gap $\Delta(V_{\rm g})<\Delta_{\rm{Al}}$, whereas $\tau$ represents an average transparency of all the channels present in the junction. 

We fit Eq.~\ref{df} to $\delta \omega_{\rm r}(\varphi_1)$ data for Dataset I, II and III [solid lines in Fig.~\ref{fig2}a)], extracting the junction parameters $\tau(V_1)$ and $\Delta(V_1)$ (Tab.~\ref{tabS_fit_param1}). Using $L_{\rm s}$ together with the resonator impedance and total inductance obtained from material characterization and electromagnetic simulations, we estimate $\lambda=0.028$. However, we find that our data are best described by the model when including $\lambda$ as an additional fit parameter, obtaining values ranging from 0.020 to 0.035. The single-channel model captures $\delta \omega_{\rm r}(\varphi_1)$ for Dataset I and II, while it fails to reproduce Dataset III around $\varphi_1=\pi$. We observe that generalizing Eq.~\ref{df} for a two-channel ABS model (see Supplementary) solves the discrepancy for Dataset III without producing appreciable changes for Dataset I and II [dashed lines in Fig.~\ref{fig2}a)]. 

We extract the Kerr nonlinearity from the linear dependence of the low-power induced frequency shift, $K\approx\delta \omega_{\rm kerr}/\bar{n}$, where $\bar{n}$ is the average probe photon number in the resonator, and the cubic nonlinearity from the on-off gain as $g_3\approx G/\sqrt{\bar{n}_{\rm p}}$, where $\bar{n}_{\rm p}$ is the average pump photon number in the resonator, as a function of $\varphi_1$ for the same datasets discussed above [Fig.~\ref{fig2}d)-e)]. Note that the probe and pump powers at the device are estimated from the nominal line attenuation and are therefore subject to an uncertainty of a few dB.

The Kerr nonlinearity follows a trend opposite to that of $\delta \omega_{\rm r}$, remaining weakly negative around $\varphi_1=0, 2\pi$, and reaching a maximum near $\varphi_1=\pi$. Its amplitude increases from Dataset I to II and develops a discontinuous pattern in III, following the corresponding avoided crossing in $\delta\omega_{\rm r}$. In contrast, $|g_3|$ exhibits two peaks symmetric about $\varphi=\pi$, whose amplitude and position depend on the junction transparency. No gain is observed around $\varphi_1=\pi$, over a phase range that increases from Dataset I to III, preventing the extraction of $|g_3|$ in this region. We attribute this behavior to the combination of an increasing Kerr nonlinearity and a reduced internal quality factor. 

Notably, $K$ changes sign, and can therefore be tuned to zero, by either gate or flux control. In particular, for Datasets I and II we find $K/2\pi\approx0$ kHz around $\varphi_1=0.8\pi, 1.2\pi$, where $|g_3|$ reaches approximately 35 kHz and 70 kHz, respectively.

A fourth-order expansion in $\lambda$ of the Hamiltonian of our system including the ABS pair degrees of freedom, necessary to describe $K$ and $g_3$, is beyond the scope of the present work. Instead, by combining the ABS ground state contribution calculated in Ref. \cite{Schrade}, with the dispersive term discussed in Refs.~\cite{Elliott, Tosi_2}, we approximate the nonlinear parameters as
\begin{equation}
        g_3 = \frac{\lambda^3}{6}c_3
    \label{g3}
\end{equation}
and 
\begin{equation}
        K = \frac{\lambda^4}{2}\left(c_4-\frac{5c_3^2}{3\tilde{c}_2}\right) + \frac{g_c^4}{\delta_-^3},
    \label{K}
\end{equation}
where $c_{3(4)}=d^{3(4)}U_{\rm ABS}/d\varphi^{3(4)}$ and $\tilde{c}_2=(\Phi_0/2\pi)^2/L_{\rm s}+c_2$. By using the parameters extracted from $\delta \omega_{\rm r}$, we find that Eqs.~\ref{g3} and \ref{K} are in qualitative agreement with the nonlinear data for all three datasets [solid lines in Fig.~\ref{fig2}d)-e)]. Generalizing Eqs.~\ref{g3} and \ref{K} to the two-channel case (see Supplementary) maintains the qualitative agreement for $g_3$, and for $K$ in Dataset I and II [dashed lines in Fig.~\ref{fig2}d)-e)]. For Dataset III, however, including the contribution of the second channel to $K$ worsens the agreement with the data, especially around $\varphi_1=\pi$, in contrast to the improved description of the linear response. This discrepancy may originate from additional terms arising from a full fourth-order expansion of the Hamiltonian
or from inter-channel coupling effects that becomes relevant at higher orders while remaining negligible for the linear response.

\begin{figure} [ht!]
        \begin{center}
                \includegraphics [width=1\columnwidth]{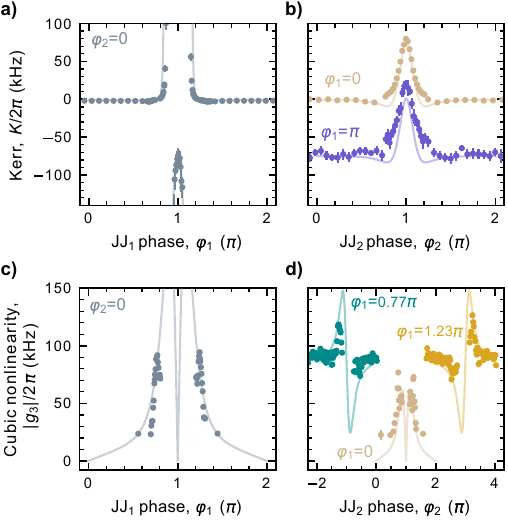}
        \end{center}
        \caption{Two-junction configuration. Kerr nonlinearity $K$ measured at gate voltages $V_{1}=1.58$ V and $V_{2}=1.94$ V, \textbf{a)} versus $\varphi_1$ with $\varphi_2=0$, and \textbf{b)} versus $\varphi_2$ with $\varphi_1=0$ (sand) and $\varphi_1=\pi$ (violet). Solid lines show $K(\varphi_1)+K(\varphi_2)$ calculated from Eq.~\ref{K}. Cubic nonlinearity $g_3$ measured at gate voltages $V_1= -0.05$ V and $V_2= 1.15$ V, \textbf{c)} versus $\varphi_1$ with $\varphi_2=0$, and \textbf{d)} versus $\varphi_2$ with $\varphi_1=0$ (sand), $\varphi_1= 0.77\pi$ (teal) and $\varphi_1=1.23\pi$ (gold). The three datasets in d) are displayed over successive phase periods for clarity. Solid lines show $|g_3(\varphi_1)+g_3(\varphi_2)|$ calculated from Eq.~\ref{g3}. Model parameters are listed in Tab.~\ref{tabS_fit_param2}.}
       
        \label{fig3}
\end{figure}

With both junctions active, we measure $K(\varphi_{2(1)})$ and $g_3(\varphi_{2(1)})$ at different $\varphi_{1(2)}$ biases [Fig.~\ref{fig3}]. At $\varphi_{1(2)}=0$, both quantities behave as in the single junction configuration, which allows the extraction of the system parameters [Tab.~\ref{tabS_fit_param2}]. On the other hand, biasing $\varphi_1=\pi$ shifts $K(\varphi_{2})$ vertically by $K(\varphi_{1}=\pi)=-70$ kHz [Fig.~\ref{fig3}a-b)]. Similarly, biasing $g_3(\varphi_1)$ to either peak shifts $g_3(\varphi_2)$ by the corresponding peak value and alters its symmetry [Fig.~\ref{fig3}c-d)]. Neglecting both the interaction between the junctions and the loop inductance, we find that the measurements are well described by the sum of the individual junction contributions. 

The double-junction configuration therefore provides an additional degree of control over the resonator nonlinearities. In particular, varying $\varphi_1$ shifts the $K(\varphi_2)=0$ operating point, although the accessible range is limited by flux noise and the steep phase dependence of $K(\varphi_1)$ near $\pi$. Furthermore, biasing JJ$_1$ close to the maxima of $|g_3(\varphi_1)|$ maintains a large cubic nonlinearity over a broad range of $\varphi_2$, enabling amplification across a wider frequency range while reducing the pump power required to reach a given gain. After maximizing $g_3$ and minimizing $K$, in the double junction configuration, we obtain an on-off gain of 25 dB and a -1 dB compression point of -138 dBm. A quantitative assessment of the amplifier performance is currently limited by the lack of a calibrated input attenuation and amplification chain (see Supplementary). Further studies will therefore be required to optimize the amplifier performance and identify its limiting factors.

In conclusion, we have characterized the linear and nonlinear response of a superconducting resonator coupled to two semiconducting Josephson junctions which we can control independently. We have shown how the Kerr nonlinearity $K$ and the cubic nonlinearity $g_3$ depend on both flux and gate voltage, finding an operating point at which $K$ is negligible and $g_3$ is sizable. To our knowledge, this constitutes the first experimental demonstration of a nonzero cubic nonlinearity in superconductor-semiconductor hybrid devices. A model based on Andreev bound state pairs shows qualitative agreement with our data. We further observed that the double-junction configuration is well described by the sum of the single-junction contributions.

No Andreev molecule phenomenology \cite{Pillet} was observed in our device, consistently with the junction separation ($\sim4~\mu$m) significantly exceeding the superconducting coherence length (hundreds of nm). However, investigating future devices with much closer junctions, as in Refs.~\cite{Haxell, Kurtossy}, would allow measurement of the nonlinearities induced by the Andreev molecule potential.

Our findings open new pathways for semiconducting Josephson junctions to be deployed in three-wave mixing-based amplifiers, quantum control of bosonic modes, and coupling schemes such as beamsplitter interaction.

\section*{Acknowledgments}

We thank V. Fatemi, M. Jirlow, T. Abad, F. J. Matute-Ca\~nadas and A. Levy Yeyati for fruitful discussions.
This work was financially supported by the European Union’s H2020 Research and Innovation Program, Grants No. 804988 (SiMS) and No. 828948 (AndQC), by the Army Research Office (ARO), Grant No. W911NF2210053, by the Novo Nordisk Foundation, project SolidQ, and by the Carlsberg Foundation. The device discussed in this work was fabricated at the Chalmers Myfab cleanroom facility.
S.G. acknowledges financial support from the European Research Council (Grant No. 101041744 ESQuAT), and from the Knut and Alice Wallenberg Foundation through the Wallenberg Centre for Quantum Technology (WACQT). External interest disclosure: S.G. is a co-founder and equity holder in Sweden Quantum AB.

\newpage

\onecolumngrid

\renewcommand{\thefigure}{S\arabic{figure}}
\setcounter{figure}{0} 
\renewcommand{\theequation}{S\arabic{equation}}
\setcounter{equation}{0} 
\renewcommand{\thetable}{S\arabic{table}}
\setcounter{table}{0} 
\newpage
\section*{Supplementary Material}

\subsection{Device}
\label{supp_sec_device}
We fabricate the devices on single-side polished, 430-$\mu$m-thick sapphire substrates cleaned using a standard SC-1 process. After heating the substrate to $660~^{\circ}\mathrm{C}$, we deposit approximately 10~nm thick NbTiN film by reactive sputtering. The resulting film has a kinetic inductance of $L_{\mathrm{k}}\approx10$ pH$/\Box$ and a critical temperature of $T_c\approx15$~K. We define alignment markers by electron-beam lithography (EBL) and lift-off of an 80~nm thick evaporated Ti/Au layer. We subsequently pattern the NbTiN film by negative EBL process using ma-N2403 resist and etch the metal by inductively coupled plasma reactive ion etching in a Cl$_2$/Ar (4/50 sccm) plasma, defining the resonator, microwave feedlines, and the coarse gate and flux-line circuitry. Additional details on the fabrication process and resonator design can be found in Ref.~\cite{Cools_sup}.

\begin{figure*} [ht!]
        \begin{center}
                \includegraphics [width=0.9\textwidth]{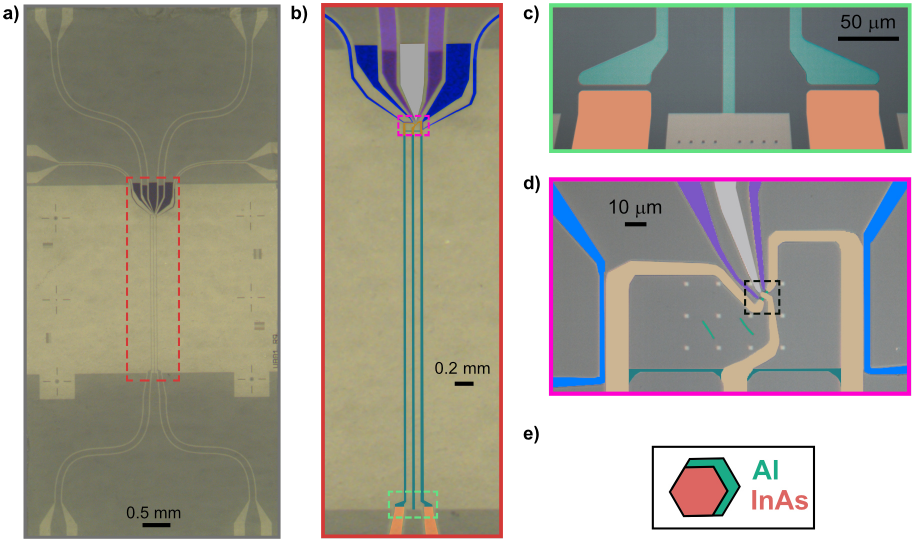}
        \end{center}
        \caption{Device. \textbf{a)} Optical micrograph of the device. The region outlined in red is magnified in b). \textbf{b)} False-colored optical micrograph showing the resonator structure (cyan), the differential feedlines (orange), the gate lines (purple) with their ground-shielding line (grey), the flux lines (blue) and the loops embedding the nanowire near the resonator voltage node (yellow). The ground plane and the substrate are left uncolored. The regions outlined in green and pink are magnified in c) and d), respectively. \textbf{c)} False-colored optical micrograph showing the coupling capacitors between the resonator and the feedlines. The color code is the same as panel in b). \textbf{d)} False-colored optical micrograph of the nanowire region located near the resonator voltage node. The nanowires are shown in green, while the rest of the circuitry follows the same color code as in b). The region enclosed by the black rectangle is shown in the SEM image in Fig.~\ref{fig1}a). \textbf{e)}: Schematic cross-section of the nanowire. The Al shell is 27~nm thick and the total diameter is approximately 180~nm. }
        \label{figS1}
\end{figure*}

We dice the wafer into $10\times5~\mathrm{mm}^2$ chips. We then deposit several hybrid Al/InAs nanowires near the resonator voltage node using a nanomanipulator. The weak links are defined by selectively removing the Al shell using an MF319 wet etch. The etch windows are patterned by standard positive EBL process. Following scan electron microscopy (SEM) inspection, we select the most promising nanowire on each chip for the final step of the fabrication. We fabricate the fine circuitry, including the ohmic contacts connecting the nanowire to the resonator as well as the gate and flux lines, by lift-off of a 250~nm thick evaporated Ti/Al layer. Immediately before metal deposition, a 30~s Ar ion milling step is performed to remove the native oxide from the epitaxial Al shell and improve the contact transparency between the nanowire and the evaporated Ti/Al contacts. The chip is then ready for measurements [see Fig.~\ref{figS1}].

\subsection{Experimental Setup}
\label{S_sec_setup}
\begin{figure*} [ht!]
        \begin{center}
                \includegraphics [width=0.8\textwidth]{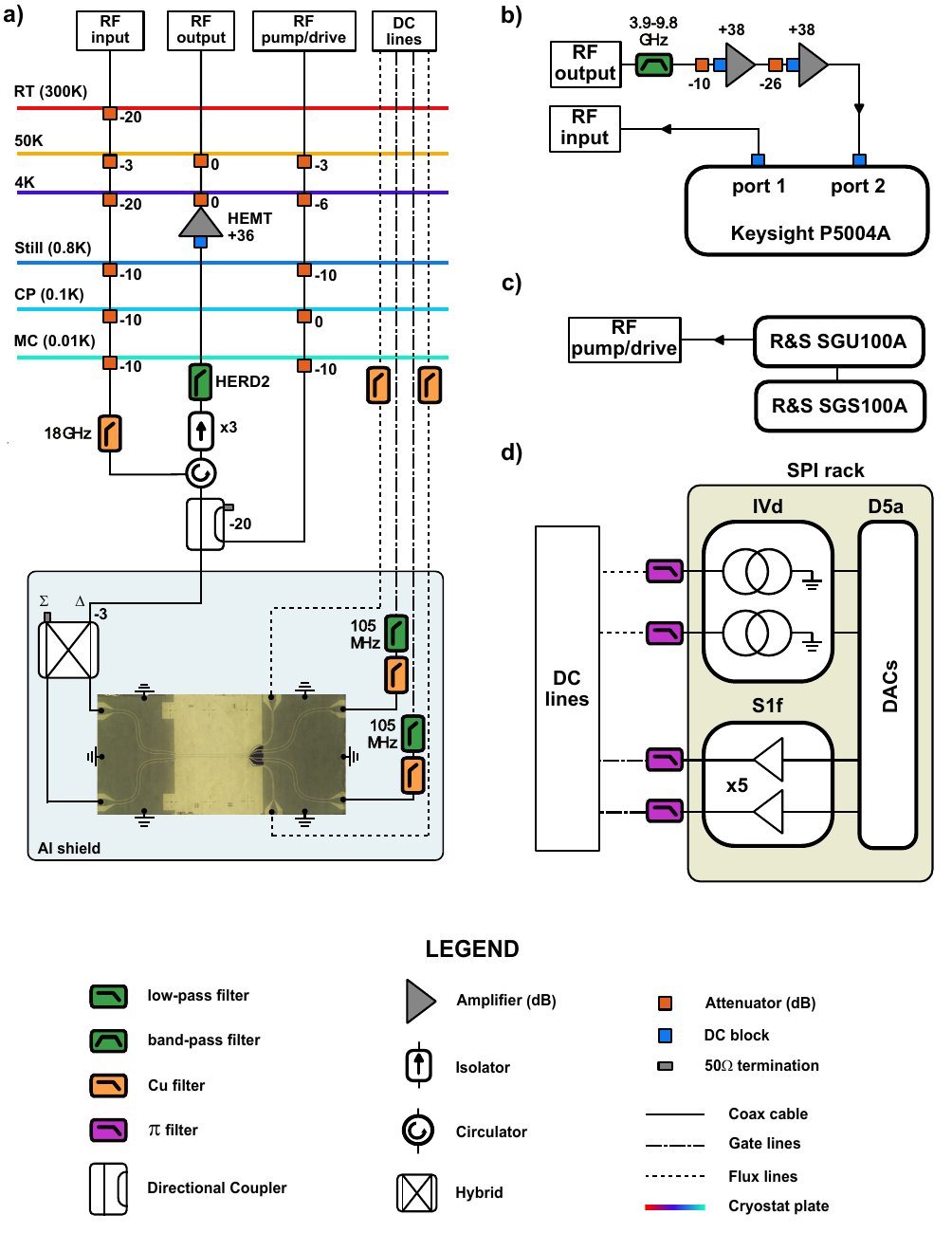}
        \end{center}
        \caption{Experimental setup. \textbf{a)} Cryostat schematic. \textbf{b)} Room temperature setup for the resonator readout. \textbf{c)} Room temperature setup for the pump/drive signal source. \textbf{d)} Room temperature setup for the gate (D5a + S1f modules) and flux (D5a + IVd modules) signal generation. The DC lines are twisted pairs from the $\pi$-filter to the mixing chamber plate. The flux lines are superconducting below the 4K plate.}
        \label{figS2}
\end{figure*}

We perform the experiment in a Bluefors dilution cryostat with 10 mK base temperature [see Fig.~\ref{figS2}a)]. The resonator is differentially driven, and readout in reflection. The Cu powder filters are fabricated in-house and, unless otherwise specified in the schematic, have a cut-off frequency of approximately 4 GHz. The room temperature setup used for the readout of the resonator [see Fig.~\ref{figS2}b)] consists of a vector network analyzer (VNA, Keysight P5004A), and additional filtering and amplification stages for the output line. On the other hand, the pump/drive line control [see Fig.~\ref{figS2}c)] consists of a signal source generator (Rohde \& Schwarz SGS100A) equipped with an upconverter (Rohde \& Schwarz SGU100A).

Specification about the modular control system SPI-rack used for the gate and flux lines [see Fig.~\ref{figS2}d)] can be found in Ref.~\cite{spi_sup}. The $\pi$ filters at room temperature absorb noise in the range 10~MHz-10~GHz. Specification for the infrared-blocking filter HERD2 can be found in Ref.~\cite{Andersson_sup}.

\newpage

\subsection{Extended gate and flux spectroscopy}
\label{S_sec_gate_and_flux}
Flux and gate lines crosstalk is compensated by applying the following linear transformation:
\begin{equation}
\begin{aligned}
\binom{I_1}{I_2} &=
\begin{pmatrix}
1 & -0.11 \\
-0.22 & 1
\end{pmatrix}
\binom{I^0_1}{I^0_2}
\qquad
\rm{and}
\qquad
\binom{V_1}{V_2} &=
\begin{pmatrix}
1 & -0.042 \\
-0.03 & 1
\end{pmatrix}
\binom{V^0_1}{V^0_2},
\end{aligned}
\end{equation}
where $I_i(V_i)$ and $I^0_i(V^0_i)$ are the actual and effective current(voltage) values sent to the device, respectively. 
\begin{figure*} [ht!]
        \begin{center}
                \includegraphics [width=0.8\textwidth]{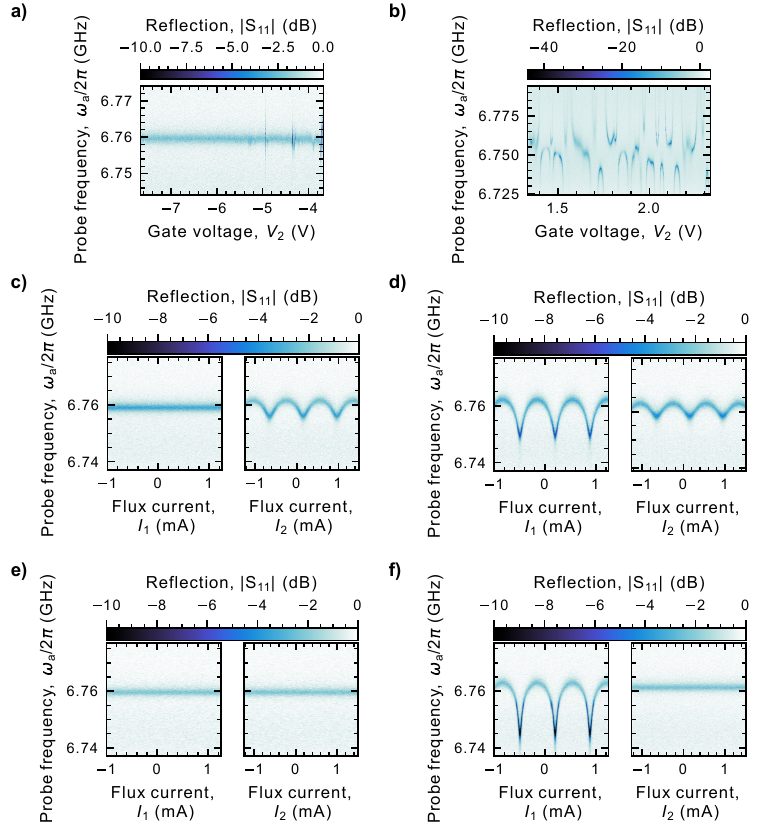}
        \end{center}
        \caption{Gate and Flux spectroscopy. \textbf{a)} Single-tone spectroscopy as a function of $V_2$ with JJ$_1$ piched-off, showing that JJ$_2$ enters in a pinch-off region for $V_2<-7$V. \textbf{b)} Single-tone spectroscopy as a function of $V_2$ in the active junction regime with $\varphi_2=\pi$, showing several avoided crossing patterns related to tuning of the conduction channels. \textbf{c-d-e-f)} Single-tone spectroscopy as a function of $I_1$ (left panels) and $I_2$ (right panels) for four different gate regimes: single-junction regime [JJ$_1$ off(on) and JJ$_2$ on(off) in c(f)), respectively], double-junction regime in d) and both junctions pinched off in e). }
        \label{figS_gate_flux}
\end{figure*}

The flux and gate crosstalk compensation enable independent junction control [see Fig.~\ref{fig1}c) and Fig.~\ref{figS_gate_flux}]. Following results for high-impedance resonator in  Ref.~\cite{Shvetsov_sup}, we neglect the loop inductance, so that the flux current $I_{\rm i}$ is converted in superconducting phase across the junction $\varphi_{\rm i}$ by linear transformation, $\varphi_{\rm i}= 2\pi(I_{\rm i} - \delta_{\rm i})/\vartheta_{\rm i}$, where $\delta_{\rm i}$ is the current corresponding to a minimum of the $f_{\rm r}(I_{\rm i})$ oscillation and $\vartheta_{\rm 1(2)}=0.138(0.158)$ mA is the oscillation period.

\subsection{Nonlinear data extraction}
\label{S_sec_nonlinearities}

\begin{figure*} [ht!]
        \begin{center}
                \includegraphics [width=1\textwidth]{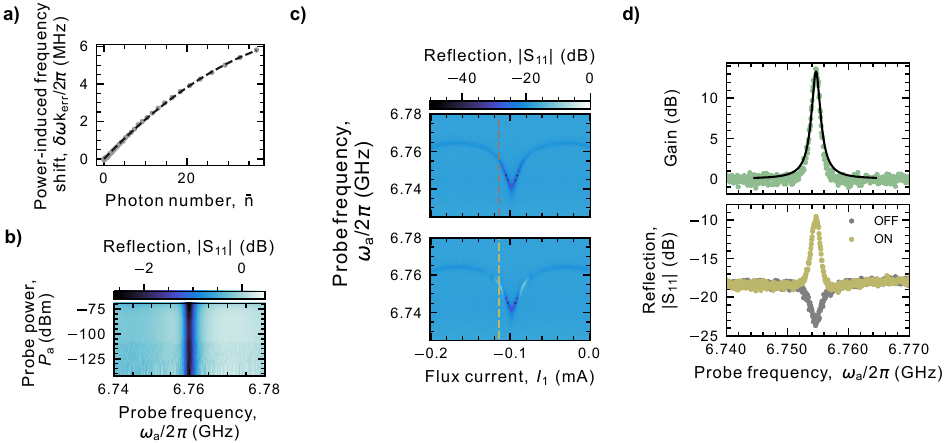}
        \end{center}
        \caption{Nonlinear data extraction. \textbf{a)} Power-induced frequency shift, $\delta\omega_{\rm kerr}$, as a function of the average probe photon number in the resonator, $\bar{n}$, for the same dataset shown in Fig.~\ref{fig1} e). The dashed black line represents the function $\delta\omega_{\rm kerr}=2K\bar{n} +K'\bar{n}^2$ fitted to the data. Fitting parameters are $K/2\pi=230$ kHz and $K'/2\pi=2$ kHz. \textbf{b)} Single-tone spectroscopy as a function of probe power $P_{\rm a}$ with both junction pinched-off. \textbf{c)} Single-tone spectroscopy as a function of flux current $I_1$ without (top) and with (bottom) a pump tone at $\omega_{\rm p}=2\omega_{\rm r}$. The gate configuration is the same as Dataset II in Fig.~\ref{fig2}. The dashed vertical lines highlight the traces shown in panel d). \textbf{d)} Line cuts from panel c). \textit{Bottom}: low-power resonator traces measured with (yellow) and without (gray) the pump tone at $\omega_{\rm p}/2\pi=13.518$ GHz. \textit{Top}: gain profile (green) obtained from the difference between the ON and OFF reflection traces shown below. The black line is a Lorentzian fit (Eq.~\ref{S_lore_gain}) from which the on-off gain $G_0$ is extracted.}
        \label{figS_kerr_g3}
\end{figure*}

The resonator Kerr $K$ is related to the power-induced frequency shift $\delta \omega_{\rm{kerr}}$ as a function of the average photon number $\bar{n}$ at the resonant frequency $\omega_{\rm{r}}$ in the cavity. We convert the probe power at the resonator, $P_{\rm a}$, estimated from the VNA output, the nominal input line attenuation and an additional -6~dB attenuation accounting for cables losses, into $\bar{n}$ using \cite{Cools_sup}:
\begin{equation}
    \bar{n}=\frac{4 Q_{\rm l}^2}{Q_{\rm c}}\frac{P_{\rm a}}{2\pi \hbar\omega_{\rm{r}}}\frac{Z_0}{Z_{\rm r}},
    \label{S_eq_photon_number}
\end{equation}
 where $Q_{\rm l}$ is the loaded quality factor of the resonator, $Z_0=50~\Omega$ is the feedline impedance, $Z_{\rm r}=290~\Omega$ is the characteristic impedance of the resonator as extracted from simulation and $h$ is the Plank constant. For sufficient low $\bar{n}$, $\delta \omega_{\rm{kerr}}(\bar{n})$ shows a quadratic behavior [see Fig.~\ref{figS_kerr_g3}a)] that can be described by the following expression \cite{Frattini_sup}:
\begin{equation}
   \delta \omega_{\rm{kerr}}=2K\bar{n}+K'\bar{n}^2,
    \label{S_eq_k}
\end{equation}
where $K$ represents the Kerr of the resonator. A power sweep of the bare resonator (i.e., with both junctions pinched-off) shows no power-induced frequency shift in the considered range of $P_{\rm a}$ [see Fig.~\ref{figS_kerr_g3}b)], from which we conclude that the Kerr data do not include any effects due to the kinetic inductance of the NbTiN or Al.

To measure the three-wave mixing gain as a function of $\varphi$, we perform single-tone spectroscopy as a function of flux current with and without a pump tone applied at $\omega_{\rm p}(\varphi)=2\omega_{\rm r}(\varphi)$ [Fig.~\ref{figS_kerr_g3}c)] from which we can extract ON and OFF traces, whose difference in amplitude gives the gain profile $G(\omega)$ [Fig.~\ref{figS_kerr_g3}d)]. For each flux bias, the on-off gain $G_0$ is approximated by the amplitude of a Lorentzian function fitted to the gain profile in linear scale and around its maximum: 
\begin{equation}
    G(\omega_{\rm {a}}) =1+\frac{G_0}{1 + \left(\frac{\omega_{\rm{a}}-\omega_{0}}{\Gamma}\right)^2},
    \label{S_lore_gain}
\end{equation}
where $\omega_0$ is the frequency at which the gain is maximum and $\Gamma$ is the half-width at half-maximum.
The cubic nonlinearity $g_3$ is then extracted from $G_0$ by solving \cite{Sivak_sup}
\begin{equation}
    G_0 = 1 + \frac{4 \kappa^2|2g_3\sqrt{\bar{n}_{\rm p}}|^2}{\left(\delta^2_{\rm p} + \kappa^2/4 - 4|2g_3\sqrt{\bar{n}_{\rm p}}|^2\right)^2},
    \label{S_eq_g3_gain}
\end{equation}
where $2\pi\kappa=\omega_{\rm r}/Q_{\rm l}$, $\delta_{\rm p}=\omega_{\rm p}-2\omega_{\rm r}$ and $\bar{n}_{\rm p}$ is the intracavity photon number at the pump frequency given by a generalization of Eq.~\ref{S_eq_photon_number} \cite{Aspelmeyer_sup}:
\begin{equation}
    \bar{n}_{\rm p}=\frac{\kappa_{\rm e}}{\left(\omega_{\rm p}-\omega_{\rm r}\right)^2+\left(\kappa/2\right)^2}\frac{P_{\rm p}}{\hbar \omega_{\rm p}},
    \label{S_eq_np}
\end{equation}
 with $2\pi\kappa_{\rm e}=\omega_{\rm r}/Q_{\rm c}$, and $P_{\rm p}$ pump power at the resonator estimated from the signal generator output, nominal pump-line attenuation and an additional -6~dB attenuation accounting for the cables losses \cite{Cools_sup}.

This method has been used to extract $G_0$ and then $g_3$ both in single [Figs.~\ref{fig2}e),~\ref{figS_JJ2}] and double [Figs.~\ref{fig3}c,d)] junction configuration. In these cases, $P_{\rm p}$ was chosen to get a around 10 dB amplification or lower, in order to avoid large-gain correction in Eq.~\ref{S_eq_g3_gain}.

\begin{figure*} [ht!]
        \begin{center}
                \includegraphics [width=0.6\textwidth]{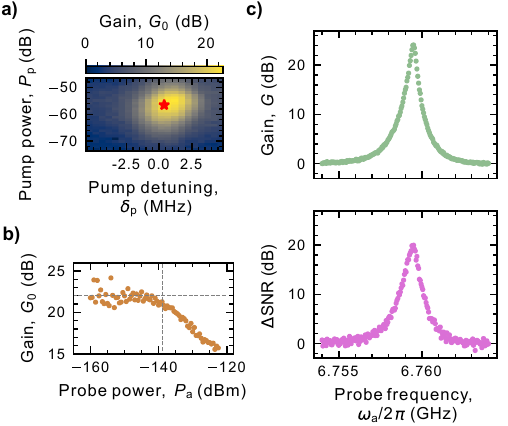}
        \end{center}
        \caption{Amplification performance. \textbf{a)} Heat-map showing the on-off amplification gain $G_0$ versus pump power $P_{\rm p}$ and pump detuning $\delta_p= \omega_{\rm p}-2\omega_{\rm r}$. Gate voltages are $V_1=0.25$ V and $V_2=0.5$ V. Phase biases are $\varphi_1\approx0.8\pi$ and $\varphi_2\approx0.75\pi$. The red stars marks the operating point used in b) and c).  \textbf{b)} Gain $G_0$ versus probe power $P_{\rm a}$. The dashed lines highlight the -1dB compression point at -138 dBm.  \textbf{c)}  On-off amplification gain $G$ (top) and corresponding signal-to-noise-ration improvement $\Delta SNR$ (bottom) as a function of probe frequency $\omega_{\rm a}$. The on-off gain shown in a,b) is extracted following the procedure described in Sec.~\ref{S_sec_nonlinearities}, whereas the data discussed in panel c) are obtained as described in Sec.~\ref{S_sec_JPA}.}
        \label{fig4}
\end{figure*}

\subsection{Parametric amplifier performance}
\label{S_sec_JPA}

We characterize the parametric amplification in the double-junction configuration. After choosing gate voltages and flux biases in order to minimize $K$ and maximize $g_3$, we sweep the pump power $P_{\rm p}$ and detuning $\delta_{\rm p}$ for maximum gain [Fig.~\ref{fig4}a)]. Using these optimized parameters, we characterize the amplifier dynamic range by measuring the maximum gain as a function of the probe power $P_{\rm a}$, obtaining a $-$1-dB compression point of -138 dBm [Fig.~\ref{fig4}b)].

With the optimized pump parameters and for probe powers below the compression point, we obtain a maximum on-off gain of 25 dB with a bandwidth of approximately 2 MHz and a signal-to-noise improvement of $\Delta \rm{SNR}=20$ dB [Fig.~\ref{fig4}b)]. In this case, the amplitude $|S^{\rm{ON}(OFF)}_{11}|$ and noise $\langle S^{\rm{ON}(OFF)}_{11}\rangle$ data are obtained at each frequency as average and variance over 400 data-points measured with the VNA in continuous-wave mode, respectively. Gain and $\Delta \rm{SNR}$ are then obtained as $G=|S^{\rm{ON}}_{11}|/|S^{\rm{OFF}}_{11}|$ and 
$\Delta \rm{SNR}=(|S^{\rm{ON}}_{11}|^2\langle S^{\rm{OFF}}_{11}\rangle)/(|S^{\rm{OFF}}_{11}|^2\langle S^{\rm{ON}}_{11}\rangle)$.

The fact that $G_0$ decreases and eventually disappears for large pump detuning [Fig.~\ref{fig4}a)], i.e. when the three-wave-mixing condition $\omega_{\rm p}=2\omega_{\rm r}$ is not satisfied, excludes that our results are dominated by two-level systems saturation.

\subsection{Effective Andreev bound state pairs model}
\label{S_sec_ABS}

\begin{figure*} [ht!]
        \begin{center}
                \includegraphics [width=0.8\textwidth]{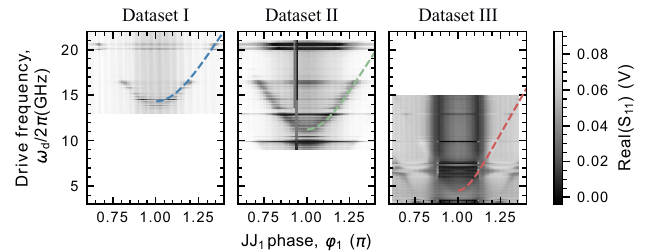}
        \end{center}
        \caption{Two tone spectroscopy corresponding to Dataset I (left), II (center) and III (right) in Fig.~\ref{fig2}a), where the dashed line represent $\hbar\omega_{\rm ABS}=2\Delta\sqrt{1-\tau\sin^2\left(\varphi/2\right)}$ with $\tau=(0.8,0.86,0.97)$ and $\Delta=(16, 15, 13)$~GHz for Datasets I, II and III.}
        \label{figS_2tone_JJ1}
\end{figure*}

From a two-tone spectroscopy around $\varphi_1=\pi$ [see Fig.~\ref{figS_2tone_JJ1}] we get a first estimation of $\tau$ and $\Delta$ for the gate configuration explored in the main text for JJ$_1$ [Fig.~\ref{fig2}a)]. These values are then used as initial guesses when fitting Eq.~\ref{df} (main text) to the $\delta\omega_{\rm r}$ data. In particular, the fitting parameters are $\tau$ and $\lambda$, while $\Delta$ is kept fixed. 

The parameter $\lambda$ is the product between the participation ratio $p=L_{\rm s}/(L_{\rm r}+2L_{\rm s})$ and the reduced zero-point-fluctuation $2\pi\Phi_{\rm zpf}/\Phi_0$ \cite{Park_sup}, where $\Phi_{\rm{zpf}}=\sqrt{\hbar Z_{\rm r}/2}$ is the resonator zero-point fluctuation, $Z_{\rm r}=290~\Omega$ is the characteristic impedance of the resonator as extracted from simulation, $\hbar$ is the reduced Plank constant and $\Phi_0$ is the flux quantum.

We can generalize the phenomenological ABS model taking into account a second channel by following Ref.~\cite{Metzger_sup}. Considering two channels with transparency $\tau_1$ and $\tau_2$, the effective ABS potential reads 
\begin{equation}
    U_{\rm{2ABS}}=\Delta_{\rm eff}\left(\sqrt{1-\tau_1\sin^2\left(\frac{\varphi}{2}\right)}+ \sqrt{1-\tau_2\sin^2\left(\frac{\varphi}{2}\right)}\right)=U^{(1)}_{\rm ABS}+U^{(2)}_{\rm ABS}.
\end{equation}
The induced frequency shift then becomes (as in the main text, $\hbar=1$):
\begin{equation}
        \delta \omega_r = \sum_{m=1}^{2}\left(-\lambda^2c^{(m)}_2 + \left(g^{(m)}_{\rm c}\right)^2\left(\frac{2}{\omega^{(m)}_{\rm{ABS}}}-\frac{1}{\delta^{(m)}_-}-\frac{1}{\delta^{(m)}_{+}}\right)\right),
    \label{df_2ch}
\end{equation}
where $c^{(m)}_2=d^2U^{(m)}_{\rm{ABS}}/d\varphi^2$, $\omega^{(m)}_{\rm ABS}=2U^{(m)}_{\rm ABS}$,  $\delta^{(m)}_{\pm}=\omega^{(m)}_{\rm{ABS}}\pm\omega^0_{\rm{r}}$ and $g^{(m)}_{\rm c}=\lambda\sqrt{1-\tau_{m}}\left(dU^{(m)}_{\rm{ABS}}/d\varphi\right)\tan\left(\varphi/2\right)$. We note that in Eq.~\ref{df_2ch} we consider the two channels to be uncoupled. 

Similarly, we generalize the cubic and Kerr nonlinearity:
\begin{equation}
        g_3 = \frac{\lambda^3}{6}\tilde{c}_3
    \label{g3_2ch}
\end{equation}
and 
\begin{equation}
        K = \frac{\lambda^4}{2}\left(\tilde{c}_4-\frac{5\tilde{c}_3^2}{3\tilde{c}_2}\right) +  \sum_{m=1}^{2}\frac{\left(g^{(m)}_c\right)^4}{\left(\delta^{(m)}_-\right)^3},
    \label{K_2ch}
\end{equation}
where $\tilde{c}_n=d^n\tilde{U}_{\rm ABS}/d\varphi^n$, with
\begin{equation}
    \tilde{U}_{\rm ABS} = \left(\frac{\Phi_0}{2\pi}\right)^2\frac{\left(\varphi-\varphi_{\rm{ext}}\right)^2}{2L_{\rm loop}}-U^{(1)}_{\rm ABS}-U^{(2)}_{\rm ABS}. 
\end{equation}

As for the double junctions configuration, Eqs.~\ref{K} and \ref{g3} (main text) are fitted to the $K$ and $|g_3|$ data for JJ$_1$(JJ$_2$) while $\varphi_2(\varphi_1)=0$ [Fig.~\ref{fig3}]. Obtained parameters are listed in Tab.~\ref{tabS_fit_param2}.

\begin{table}[h]
\centering
\begin{tabular}{|c|c|c|c|c|c|}
\hline
 Junction & Dataset & $\Delta$ (GHz) & $\tau$ & $\tau_2$ & $\lambda$  \\
\hline
\multirow{6}{*}{JJ$_1$}
& \multirow{2}{*}{I}
    & 16 & 0.739 & - & 0.032 \\ \cline{3-6}
&
    & 16 & 0.797 & 0.379 & 0.026 \\ \cline{2-6}

& \multirow{2}{*}{II}
    & 15 & 0.809 & - & 0.035 \\ \cline{3-6}
&
    & 15 & 0.851 & 0.483 & 0.028 \\ \cline{2-6}

& \multirow{2}{*}{III}
    & 9 & 0.925 & - & 0.035 \\ \cline{3-6}
&
    & 13 & 0.975 & 0.850 & 0.020 \\ \hline
    
\multirow{6}{*}{JJ$_2$}
& \multirow{2}{*}{IV}
    & 18 & 0.736 & - & 0.024 \\ \cline{3-6}
&
    & 18 & 0.752 & 0.071 & 0.023 \\ \cline{2-6}

& \multirow{2}{*}{V}
    & 15 & 0.863 & - & 0.029 \\ \cline{3-6}
&
    & 13 & 0.865 & 0.328 & 0.027 \\ \cline{2-6}

& \multirow{2}{*}{VI}
    & 9.5 & 0.900 & - & 0.035 \\ \cline{3-6}
&
    & 12.8 & 0.955 & 0.400 & 0.025 \\ \hline
\end{tabular}
\caption{Parameter describing single-junction configuration data for JJ$_1$ [Fig.~\ref{fig2}] and JJ$_2$ [Fig.~\ref{figS_JJ2}]. For each Dataset we list both the one-channel and two-channel parameters.}
\label{tabS_fit_param1}
\end{table}

\begin{figure*} [ht!]
        \begin{center}
                \includegraphics [width=0.8\textwidth]{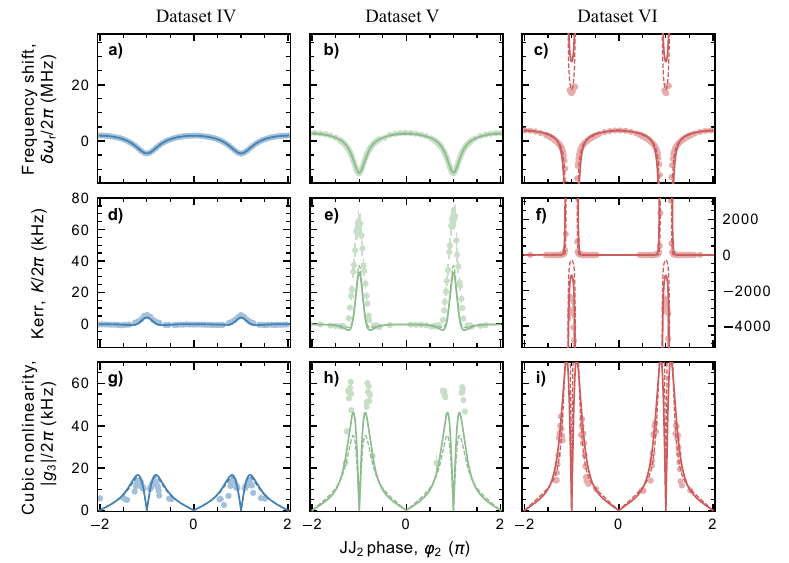}
        \end{center}
        \caption{Linear ($\delta \omega_{\rm r}$) and nonlinear ($K,~g_3$) data for the single-junction regime where JJ$_1$ is pinched-off and JJ$_2$ is active measured versus $\varphi_2$, for Dataset IV (left, $V_2=-0.55$ V), V (center, $V_2=0.65$ V) and VI (right, $V_2=1.8$ V). Solid and dashed lines represent one-channel and two-channel model, respectively. Parameters are listed in Tab.~\ref{tabS_fit_param1}. }
        \label{figS_JJ2}
\end{figure*}

\begin{table}[h]
\centering
\begin{tabular}{|c|c|c|c|c|}
\hline
 Dataset & Junction & $\Delta$ (GHz) & $\tau$ & $\lambda$  \\
\hline
\multirow{2}{*}{$K$}
& JJ$_1$ & 16 & 0.88 & 0.033 \\ \cline{2-5}
& JJ$_2$ & 10 & 0.92 & 0.023 \\ \cline{2-5}
\hline
\multirow{2}{*}{$g_3$}
& JJ$_1$ & 17 & 0.87 & 0.03 \\ \cline{2-5}
& JJ$_2$ & 15 & 0.93 & 0.038 \\ \cline{2-5}
\hline
\end{tabular}
\caption{Parameter describing double-junction configuration data [see Fig.~\ref{fig3}].}
\label{tabS_fit_param2}
\end{table}

\end{document}